\documentclass[slac_one]{revtex4}
\usepackage{graphicx}
\usepackage{fancyhdr}
\newcommand{\um}{\ensuremath{\mu\mathrm{m}}}
\newcommand{\mm}{\ensuremath{\mathrm{mm}}}
\newcommand{\cm}{\ensuremath{\mathrm{cm}}}
\newcommand{\m}{\ensuremath{\mathrm{m}}}

\newcommand{\kA}{\ensuremath{\mathrm{k\kern -0.1em A}}}
\newcommand{\uA}{\ensuremath{\mu\kern -0.1em \mathrm{A}}}

\newcommand{\kV}{\ensuremath{\mathrm{k\kern -0.1em V}}}

\newcommand{\Hz}{\ensuremath{\mathrm{Hz}}}
\newcommand{\kHz}{\ensuremath{\mathrm{k\kern -0.1em Hz}}}

\newcommand{\ns}{\ensuremath{\mathrm{ns}}}
\newcommand{\s}{\ensuremath{\mathrm{s}}}

\newcommand{\GeVc}
	   {\ensuremath{\mathrm{Ge\kern -0.1em V}\kern -0.1em /\mathrm{c}}}
\newcommand{\TeVc}
	   {\ensuremath{\mathrm{Te\kern -0.1em V}\kern -0.1em /\mathrm{c}}}

\newcommand{\GeV}{\ensuremath{\mathrm{Ge\kern -0.1em V}}}
\newcommand{\TeV}{\ensuremath{\mathrm{Te\kern -0.1em V}}}

\begin{document}

%Title of paper
\title{Integration and commissioning of the ATLAS Muon spectrometer} 
%% Paper title goes here

\author{A. Belloni, for the ATLAS Collaboration}
\affiliation{Harvard University, Cambridge, MA 02138, USA}

\begin{abstract}
The ATLAS experiment at the Large Hadron Collider (LHC) at CERN is
currently waiting to record the first collision data in spring 2009. Its
muon spectrometer is designed to achieve a momentum resolution of $10\%$
$p_T(\mu) = 1~\TeVc$. The spectrometer consists of a system of three
superconducting air-core toroid magnets and is instrumented with three
layers of Monitored Drift Tube chambers (Cathode Strip Chambers in the
extreme forward region) as precision detectors. Resistive Plate Chambers in
the barrel and Thin Gap Chambers in the endcap regions provide a fast
trigger system.

The spectrometer passed important milestones in the last year. The most
notable milestone was the installation of the inner layer of endcap muon
chambers, which constituted the last big piece of the ATLAS detector to be
lowered in the ATLAS cavern. In addition, during the last two years most of
the muon detectors were commissioned with cosmic rays while being assembled
in the underground experimental cavern. We will report on our experience
with the precision and trigger chambers, the optical spectrometer alignment
system, the level-1 trigger, and the ATLAS data acquisition system. Results
of the global performance of the muon system from data with magnetic field
will also be presented.
\end{abstract}

\maketitle

\thispagestyle{fancy}

% body of paper here - Use proper section commands
% References should be done using the \cite, \ref, and \label commands
% Put \label in argument of \section for cross-referencing
%\section{\label{}}

\section{INTRODUCTION}
The ATLAS detector is nearing its completion after a 15-years effort, and
the Large Hadron Collider is scheduled to start providing $pp$ collisions in
spring 2009. The initial (design) luminosity will be $10^{33}$
($10^{34}$)~$\cm^{-2}\s^{-1}$, with inter-bunch separation of $75$ 
($25$)~$\ns$.

The precision measurement of muon tracks is performed, over most of the
spectrometer acceptance, by Monitored Drift Tube (MDT) chambers. 
Three layers of MDT chambers arranged in concentric cylinders constitute the
barrel MDT system. In the forward region, MDT chambers are installed in
three wheels, with a coverage up to $|\eta| = 2.7$.
The MDT chambers are continuously monitored for misalignments and deformations
by a network of optical lines, which allow for the achievement of the design
resolution on sagitta measurements.
In the most forward part of the inner layer of the muon spectrometer,
where the background rate is expected to reach $1~\kHz/\cm^2$, Cathode Strip
Chambers (CSC) are used. Muons are triggered by three double layers of
Resistive Plate Chambers (RPC) in the barrel region ($|\eta| < 1.05$), and by
three stations of Thin Gap Chambers (TGC) in the endcap region
($1.05 < |\eta| < 2.4$).

The largest part of the muon spectrometer is embedded in a toroidal magnet
field, provided by three magnets: one barrel toroid, and two endcap toroids,
one per side. The total bending power is 2-6~Tm in the barrel region, and
1-7~Tm in the endcap regions. 

The four technologies of muon spectrometer chambers and the system of toroidal
magnets are described in detail in~\cite{:2008zz}.

\section{HARDWARE STATUS}
%{\bf Monitored Drift Tubes}
The MDTs are built from $3~\cm$-diameter aluminum tubes
equipped with a $50~\um$-diameter gold-plated Tungsten-Rhenium wire,
kept at 3080~V with respect to the tube wall. This setup provide a single-hit
resolution of $80~\um$ in the bending direction of tracks.
The most common types of chamber contain 6 to 8 layers of drift tubes,
which results in a chamber resolution around $35~\um$.

All MDT chambers have been installed in the ATLAS cavern, and commissioning
with cosmic rays is well under way.
The latest milestones were the installation
of the inner station of the endcap system, the Small Wheels, and of
the last seventy-two chambers of the outer station of the
endcap system, the Endcap Outer (EO) station.
This last step marked the completion of the MDT installation. A set of 62
extra endcap (EE) chambers has been staged for installation in 2009-2010.

The MDT system, excluding the EE chambers, consists of 1088 chambers,
for about 339,000 readout channels. 
The fraction of readout channels which are not included in the ATLAS
Trigger and Data Acquisition (TDAQ) chain is in the order of $1\%$. It
will be reduced of one order of magnitude after the interventions planned
during the winter 2009 shutdown.

Deformations and alignment of MDT chambers are actively monitored by a set of
optical rays. The alignment of chambers with respect to the rest of ATLAS
(in particular, the inner detector) is studied utilizing cosmic rays and
beam data. 
The alignment system is fully commissioned, and even if a small fraction of the
optical lines is not usable (either for a camera or laser fault, or
unexpected obstacles along a line), the redundancy of the system covers for
the missing parts. The fraction of working sensors is about $98-99\%$
($99.7\%$ in the MDT barrel). The largest fraction of
interrupted polar lines, which are used to align chambers in different stations
(inner, middle, outer), is in one of the Endcap wheels,
where $96\%$ of the polar
lines are working. The recovery of the missing lines requires an adjustment
of the position of that Endcap middle wheel, which is scheduled for the
next shutdown.
The chamber alignment fit, which is a measure of the internal consistency
of the sensor measurements, provides an indication of the sagitta
resolution that can be achieved. The fit of the Endcap chambers indicates
an expectation around $45-50~\um$, while the fit of the Barrel points at
a resolution of $300-500~\um$.
This difference is explained by the fact that the Barrel chambers could not be
surveyed as accurately as the Endcap ones. Every Endcap chamber has four
survey targets, while, due to installation constraints, only one Barrel chamber
out of six could be surveyed, and with two targets. Straight tracks collected
during runs with muon magnets off will be utilized to complete the
alignment and provide the design momentum resolution with both barrel and
endcap chambers.

Most of the chambers of the muon spectrometer are embedded in a toroidal
magnetic field produced by three superconducting magnets. The muon magnet
system has been fully commissioned, and all of its parts have been extensively
tested with full current, $20.5~\kA$. 

CSCs are
installed in the region $|\eta|> 2.0$, at a distance of about $7~\m$ from the
interaction point. They are multiwire proportional chambers, with wires
oriented in the radial direction (more precisely, all wires in a chamber are
parallel to the central wire, which points in the radial direction).
A CSC chamber can reach a resolution of $40~\um$
in the bending direction, and $5~\mm$ in the non-bending direction.

The 32 CSCs are part of the Small Wheels and were thus installed
in February 2008. They were connected
to the ATLAS services in less than two weeks. The inclusion in the
ATLAS readout chain has been so far delayed by debugging activity on the
firmware of the CSC readout drivers, which is scheduled to end by middle
October. However, CSC chambers have been successfully read out and it has
already been checked that one of the most important quality factors for
these chambers, the rate of noise hits, is well under
control~\cite{Ivo:2008}.

Trigger information in the barrel region is provided by RPCs, gaseous
parallel electrode-plate detectors. They are installed in three double layers
mounted on the MDT chambers of the middle (two double layers) and outer (one
double layer) stations. Each RPC chamber consists of two resistive plates
kept at a distance of $2~\mm$, with an electric field
of $4.9~\kV/\mm$ between them. The signal is read out via capacitive coupling
to metallic strips mounted on the outer faces of the resistive plates. 
The RPC hit resolution is $10~\mm$ in both the bending and non-bending
directions. 
At the nominal voltage, a signal with a width of about $5~\ns$ is generated
by a track, which allows the RPC to correctly identify the bunch crossing to
which the hit belongs with a high purity.

Commissioning is promptly advancing. All 16 RPC sectors have been read out
in the ATLAS TDAQ chain, and provided trigger signals.
The fraction of unrecoverable channels is $0.5\%$, out of 359,000 in 544
chambers. The installation of RPCs will be completed when the electronics
for the 62 chambers in the ATLAS feet region will be positioned.
The last, still missing, step is the tuning of the synchronization between
the RPC trigger and the LHC clock. This activity is currently under way, and
will be completed by middle October 2008.

%{\bf Thin Gap Chambers}
Thin Gap Chambers produce a trigger signal when muons are in the
$1.05<|\eta|<2.4$ region of pseudorapidity. They are multiwire proportional
chambers
with a chamber resolution of $2-6~\mm$ in the bending direction, and $3-7~\mm$
in the non-bending direction.
Signals arrive with $99\%$ probability inside a time window of $25~\ns$.
Thus the TGCs provide both a trigger signal and a measurement of the
$\phi$ coordinate which complements the precise MDT measurement in the
radial direction.

The TGC system is in very good conditions. The fraction of dead channels is
around $0.03\%$, and by September 2008 all 24 sectors (12 per side) of the
trigger chambers have been providing triggers and tracking information.
The completely installed TGC system
consists of 3588 chambers, which total 318,000 readout channels.
A delay in the shipment of power
supplies and the start of LHC commissioning, with the consequent restrictions
on working conditions in the cavern, prevented the full integration of
the TGCs in the small wheels. These chambers are not used to produce a trigger
signal, but help tracking by providing a bi-dimensional hit which complements
the MDT information.

\section{SLOW CONTROLS AND DATA ACQUISITION}
The Detector Control System (DCS) of ATLAS is implemented as a finite-state
machine, and provides a unified framework for detector operations and
monitoring~\cite{BarriusoPoy:2008zz}.
The DCS also continuously archives the run-time parameters of the detector
hardware, and can issue automatic commands whenever
required to keep the detector in safe conditions.

The user interface provided by DCS for the control of ATLAS subdetectors
and infrastructure is in a very advanced stage. In particular, the controls
for MDT, RPC, and TGC chambers have been extensively utilized during actual
data taking.
The full control chain of services for all types of muon spectrometer chambers
is remotely managed and monitored. This includes
the management of low-voltage power supplies for chamber electronics, 
high-voltage power supplies for sense and field devices, and gas quality and
flow. 
The same framework is utilized to monitor the working conditions of the
front-end electronics.
Temperature sensors, B-field probes, and alignment lasers and cameras,
installed on MDT chambers, are also included.

A single operator for subdetector is sufficient for configuring the
subdetector for data-taking and monitoring its conditions.
This is a clear indicator of the maturity of the control and monitoring system,
and allowed for the safe operations of the muon spectrometer in the ATLAS
TDAQ chain.

The RPC and TGC detectors are fully integrated in the trigger and data
acquisition system. They have been extensively utilized to provide a L1
trigger, which has been successfully synchronized with the trigger signal
from the other detectors that concur to the production of a global L1 
trigger. The RPC L1 trigger rate with cosmic rays is around $200~\Hz$, while
the TGC rate is about $20~\Hz$ per side, with an average efficiency
above $90\%$.

MDTs and  RPCs have already been tested in the TDAQ chain
up to a L1 rate of $100~\kHz$, which corresponds to the running conditions for
the ATLAS upgrade. TGCs reached $40~\kHz$, while
CSCs are currently limited at about $3~\kHz$, but 
their work on readout firmwares is producing a steady progress towards
the $100~\kHz$ goal.

\section{COSMIC-RAYS RESULTS}
The muon spectrometer has taken part of the ATLAS
``milestone weeks'' since February 2007. In these 
special days, all available ATLAS subdetectors
were configured together to collect cosmic-ray data simultaneously.
As the beginning
of LHC commissioning was approaching, in June 2008 the collaboration shifted
towards combined running of ATLAS subdetectors on a regular basis, and
combined running is currently a standard operation.
The data discussed in this section were
collected during cosmics runs in the last milestone weeks.

\begin{figure}
\includegraphics [height=6cm]{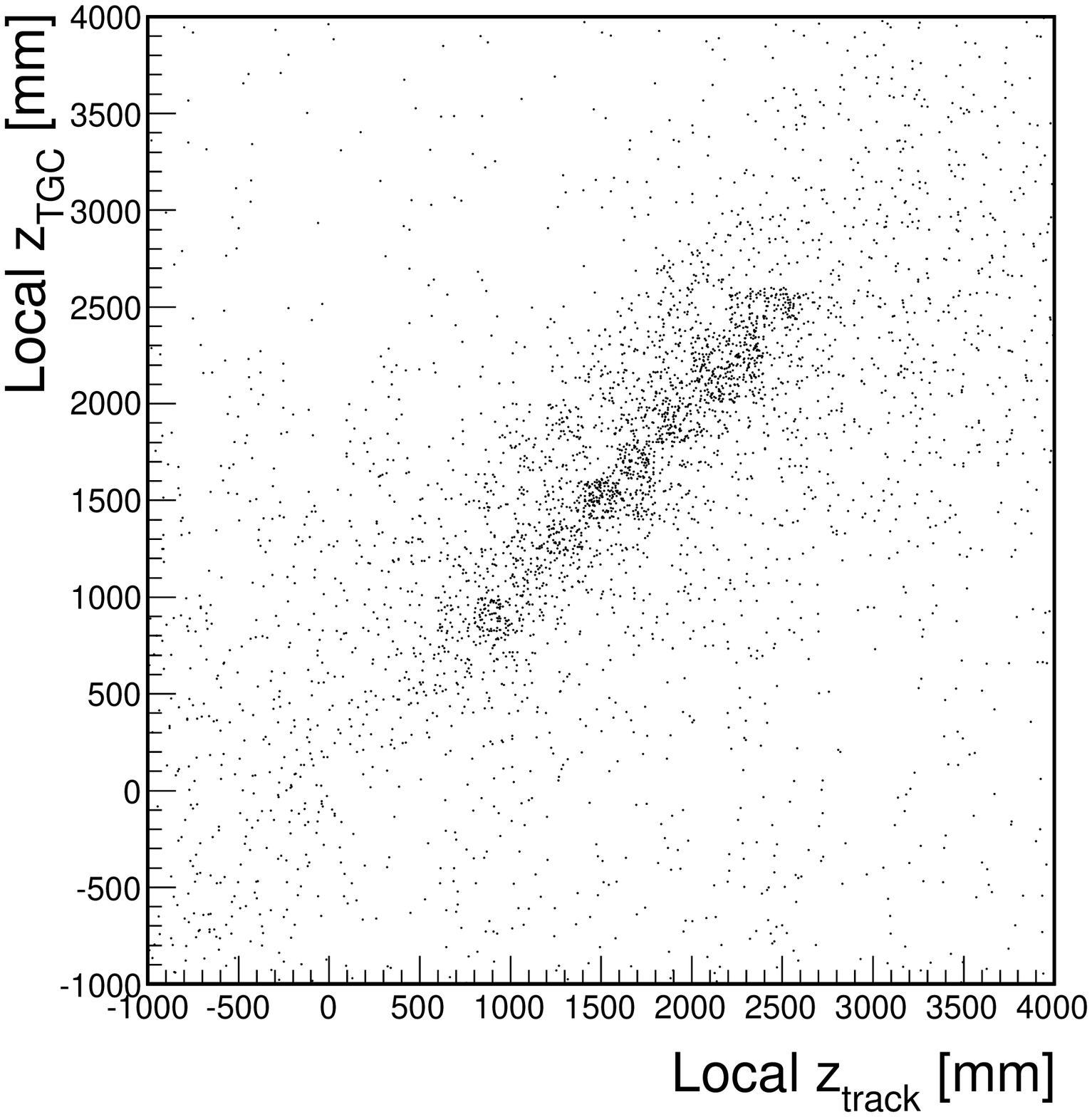}%
\includegraphics [height=6cm]{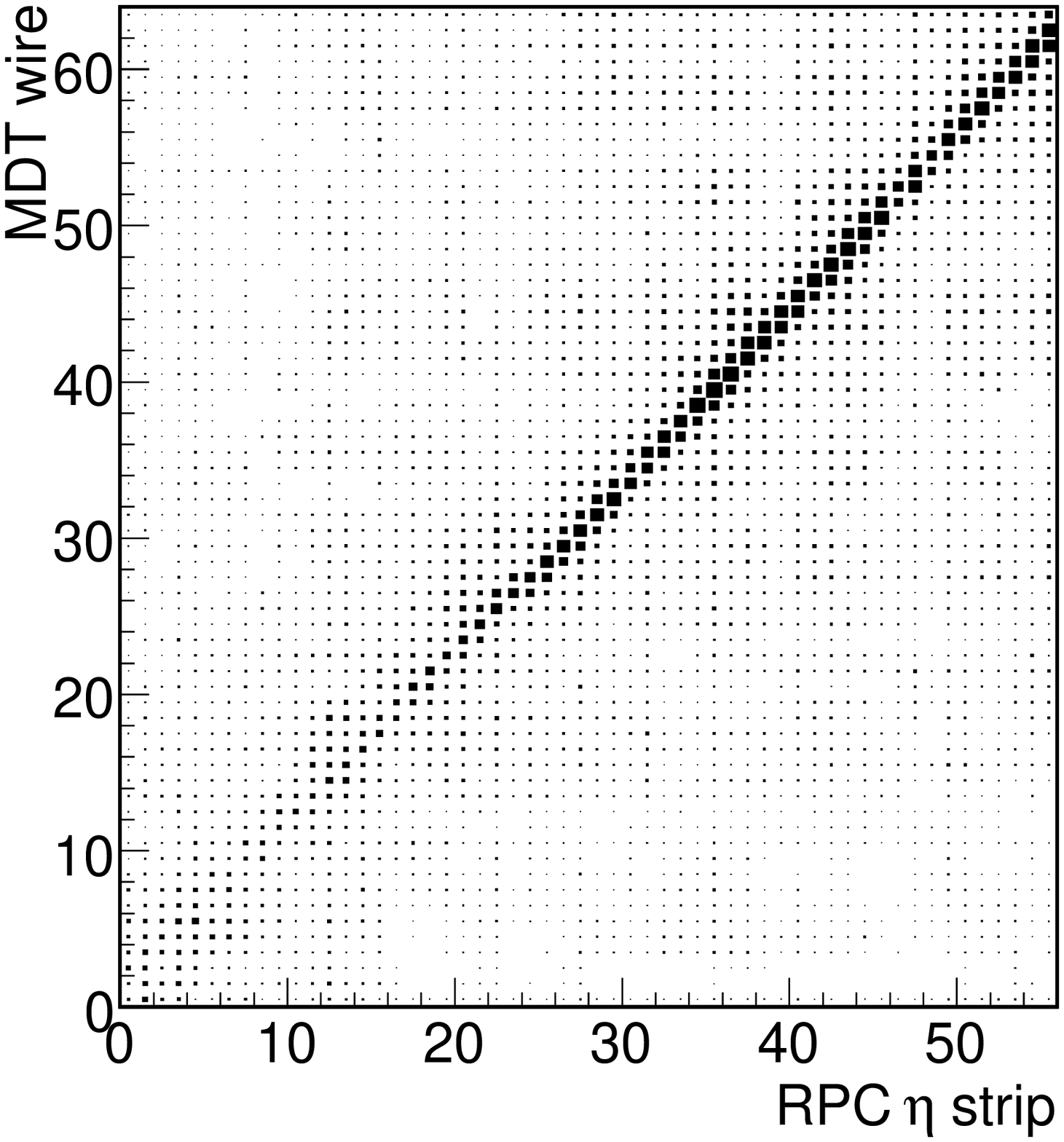}%
\includegraphics*[height=6cm]{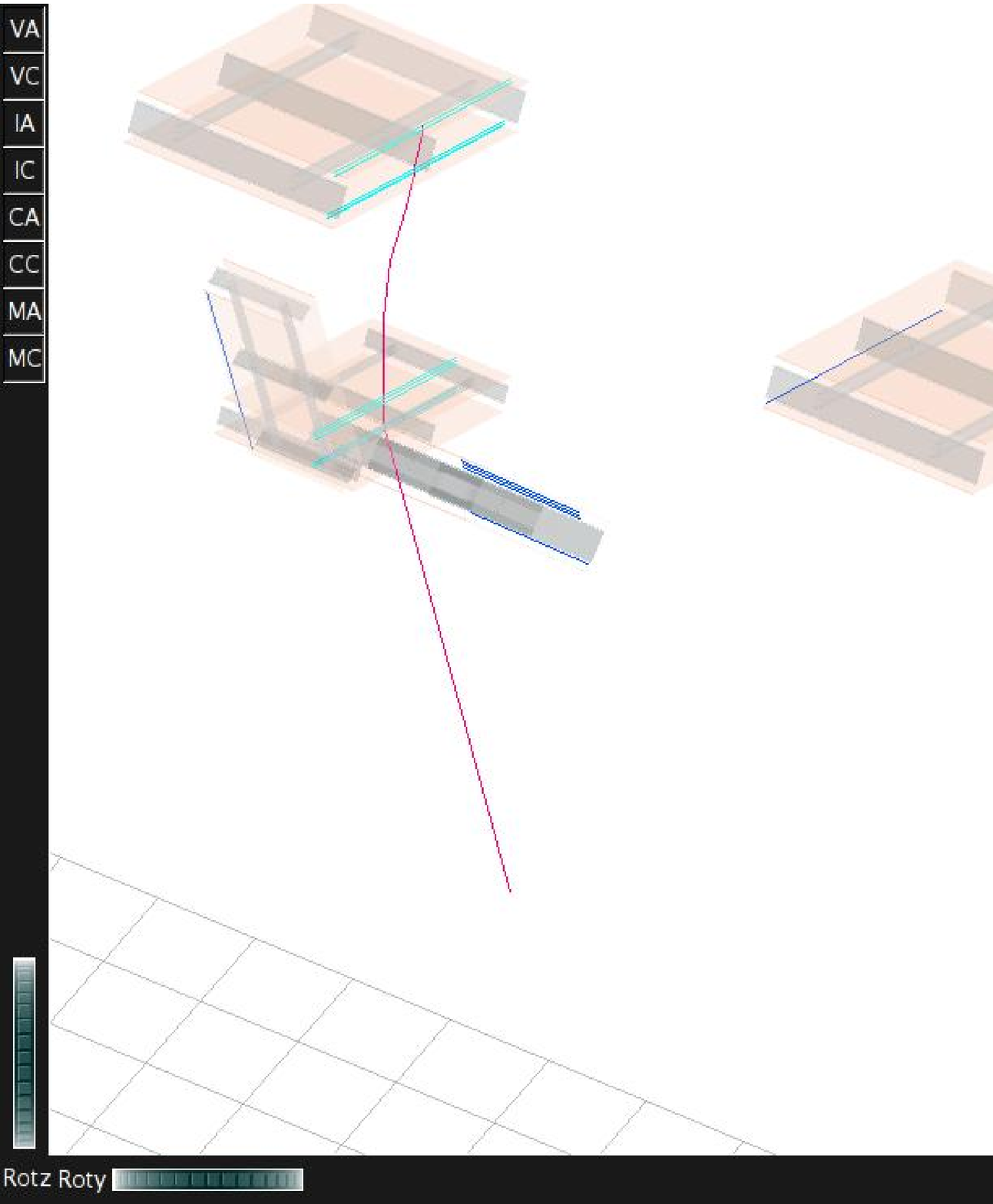}
\caption{The scatter plot on the left is produced
  by a standalone fit of muon tracks in the Endcap spectrometer. The $z$
  positions, a local coordinate in the bending direction, of TGC hits is
  compared to the extrapolated $z$ of tracks fitted using MDT hits only.
  In the center, the plot of RPC {\sl vs.} MDT channel numbers shows
  a clear correlation.
  On the right, an ATLAS event display showing a $3~\GeVc$ track in the barrel
  region.
  The track crosses three MDT chambers, one per station --- inner, middle, 
  outer --- and a pair of RPC chambers, which possibly produced the trigger
  signal.  
  \label{fig:mdtcorr}}
\end{figure}

The two scatter plots in Fig.~\ref{fig:mdtcorr} show the correlation 
between hits in the trigger and in the precision tracking chambers.
In the left plot, the hit position in $z$, a local coordinate in a TGC
chamber perpendicular to the TGC wires, is plotted versus the extrapolation
to the position of the TGC chamber of a muon segment which is fit utilizing
MDT hits only.
It is easier to show a correlation between MDT and RPC hits because the
geometry of the precision and trigger chambers overlap in the barrel region.
It is thus enough to plot the correlation between readout channel numbers 
in MDT and in RPC chambers, after a trivial correction for different mappings.
The correlation is shown in the central scatter plot in
Fig.~\ref{fig:mdtcorr} for a barrel chamber in
the MDT outer layer. Remarkably, no clean-up cuts have been applied to
produce the distribution shown.

An event display from a run taken with the barrel
toroid on at full current is presented in Fig.~\ref{fig:mdtcorr}.
The muon software reconstructs a $3~\GeVc$ track,
which crosses two RPC chambers and three MDT chambers in the barrel region.

\section{SUMMARY}
After the completion of hardware installations and connection to services,
the ATLAS muons spectrometer is quickly reaching the fully-commissioned status.
All subdetectors that compose the muon spectrometer achieved their 
design performance, for all the quantities that could be checked.
Results from the most recent cosmics runs with and 
without the toroidal field have been presented. %, and the very first
The collaboration is eagerly waiting for physics collisions, scheduled for
spring 2009.

\begin{acknowledgments}
We are greatly indebted to all CERN's departments and to the LHC project for
their immense efforts. We are grateful to all the funding agencies which
supported the construction and the commissioning of the ATLAS
detector and also provided the computing infrastructure.
\end{acknowledgments}

\end{document}